\documentclass[11pt]{JHEP}
\usepackage{latexsym}
\usepackage{amssymb,amsfonts}
\usepackage{amsfonts}
\font\mybb=msbm10 at 11pt
\def\bb#1{\hbox{\mybb#1}}

\renewcommand{\theequation}{\arabic{section}.\arabic{equation}}
\title{Action for the eleven dimensional multiple M-wave system}

\author{Igor A. Bandos
$^{\dagger\ddagger}$
\\
$^{\dagger}$
Department of
Theoretical Physics, University of the Basque Country UPV/EHU,
\\
P.O. Box 644, 48080 Bilbao, Spain
 \\
 $^{\ddagger}$
IKERBASQUE, Basque Foundation for Science, 48011, Bilbao, Spain}

\date{November 5, 2012}

\abstract{We present the covariant supersymmetric and $\kappa$--symmetric action for a system of N nearly coincident M-waves (multiple M$0$-brane system) in flat eleven dimensional superspace.}


\begin{document}

\section*{Introduction}

When speaking about M$p$-branes, eleven dimensional (11D) supersymmetric extended objects playing an important role in String/M-theory, one usually mentions M2--brane, also known as 11D supermembrane (p$=$2) \cite{BST1987} and M5-brane \cite{hs2,blnpst,schw5}. However, as it was noticed in \cite{B+T=Dpac}, one more supersymmetric 11D object exists.  This is 11D massless superparticle ($p=0$), which can be called M$0$--brane or M--wave. The last name looks natural as far as 'M' refers to M-theory and hence to 11D, while 'wave' is  an appropriate name for a massless particle moving along a light--like line in spacetime. The other name, M$0$-brane, comes from the observation that (see  \cite{B+T=Dpac}) the dimensional reduction of this 11D superparticle down to 10D produces type IIA massive superparticle which is called D$0$-brane because it belongs to the numerous family of D$p$--branes (Dirichlet $p$--branes) of the type II superstring theories (see \cite{B+T=Dpac} and refs therein).

Being the simplest representative of the family of M-branes, the M-wave provides a natural starting point for studying their properties and the toy model to attack difficult problems related to them. One of such problems is the search for an effective action for the system of multiple M2 (mM2) and multiple M5--branes (mM5).

The dimensional reduction of these hypothetical  mM$p^\prime$ actions should produce the actions for mD$p$ (multiple Dp--brane) systems. For these a (very) low energy limit is provided \cite{Witten:1995im} by the maximal $d=p+1$ dimensional supersymmetric Yang--Mills  (SYM) theory with the gauge group $U(N)$. This includes $(9-p)$ Hermitian matrices of scalar fields, $\tilde{{\bb X}}{}^{\tilde{i}}$, the diagonal elements of which describe the positions of different D$p$-branes while the off--diagonal elements account for the strings stretched between different D$p$-branes. The SYM description provided the basis for the search for the complete (a more complete) nonlinear description of mD$p$--system, similar to the description of single D$p$--brane in {\it e.g.} \cite{B+T=Dpac};
see \cite{Myers:1999ps,Dima+=JHEP03,Howe+Linstrom+Linus} and refs. therein for the progress in this direction.

For the case of mM$5$ even the question on what should be a counterpart of the very low energy approximate SYM description of mD$p$ is still obscure (see {\it e.g.} \cite{SSW} for a relevant result and references). For the case of very low energy mM$2$ system such a problem was unsolved many years, but  recently two related models were proposed  in \cite{BLG,Gustavsson:2007vu} and \cite{ABJM}.

As far as multiple M$0$--brane (mM$0$) action is considered, a purely bosonic candidate was constructed in \cite{YLozano+=0207} as  the 11D generalization of the Myers 'dielectric D$0$-brane' action in \cite{Myers:1999ps}. On the other hand, an approximate but supersymmetric and Lorentz covariant equations of motion for mM$0$--system were obtained  in \cite{mM0=PLB} in the frame of superembedding approach (see \cite{bpstv,hs2} as well as  \cite{Dima99,IB09:M-D} and refs. therein). The generalization of these equations for the case of mM$0$--system in curved 11D supergravity superspace, which describes the generalization of the M(atrix) theory \cite{Banks:1996vh} for the case of its interaction with arbitrary supergravity background, were presented in \cite{mM0=PRL} and studied in \cite{mM0=PRD}. In \cite{mM0-pp=PRD} it was shown that, when specialized for the case of 11D pp-wave superspace, these equations reproduce (in an approximation) the so--called BMN  matrix model proposed for this background by Berenstein, Maldacena and Nastase in \cite{BMN}
(see \cite{Lozano:2005kf} for the derivation of the bosonic limit of the BMN action).
This result has confirmed that the equations of \cite{mM0=PRL,mM0=PRD} describe the Matrix theory interacting with supergravity background, but its derivation has shown that, due to their superspace origin, application  of these equations  for the case of some definite, even purely bosonic supergravity backgrounds are technically complicated: it requires the lifting of the bosonic supersymmetric solution of 11D supergravity to the complete superfield solution of the 11D superspace supergravity constraints \cite{BrinkHowe80} and to use them to specify the induced geometry on the center of energy worldline.

Hence, for applications it is  desirable to find an action which reproduces the Matrix model equations of \cite{mM0=PRL} or their generalizations. In this paper we present such action for multiple M$0$--brane system, for the case of flat target 11D superspace. This action is essentially based on the spinor moving frame formalism for the 11D massless superparticle \cite{IB07:M0} (see \cite{B90,IB+AN=95} for D=4 and D=6,10 cases and \cite{IB09:M-D} for refs. on related studies) which we briefly describe in sec. 2 below.

\section{M0--brane, 11D massless superparticle}

As single M$0$--brane (M-wave) is just the D$=$11  massless superparticle, it can be described by the 11D version of the Brink--Schwarz action \cite{B+T=Dpac} $S_{BS}$ . In the first order formalism
 \begin{eqnarray}
\label{S'M0=} S_{BS} = \int_{W^1} \left( p_a \hat{E}^{a} -{e\over 2} p_ap^a d\tau  \right) \; ,
  \end{eqnarray}
where $a=0,1,...,10$ is the $SO(1,10)$ vector index, $e(\tau)$  is the Lagrange multiplier the variation of which imposes the mass shell condition $p_ap^a=0$,
$\hat{E}^{a}= {E}^a(\hat{Z})=:d\hat{Z}{}^M(\tau)  {E}_M^{a}(\hat{Z})$ is the pull--back of the bosonic supervielbein of the 11D target superspace, $E^a= E^a(Z)= dZ^ME_M^a(Z)$,  to the worldline $W^1$ parametrized by proper time $\tau$. The pull--back is obtained by substituting the bosonic and fermionic coordinate functions $\hat{Z}^M(\tau)= (\hat{x}^\mu(\tau), \hat{\theta}^{\check{\alpha}}(\tau))$ ($\mu=0,...,9,10$, $\check{\alpha}=1,...,32$) for the superspace coordinates ${Z}^M= ({x}^\mu, {\theta}^{\check{\alpha}})$, so that
\begin{eqnarray}
\label{hEa=dthEt}
\hat{E}^{a}= d\tau  \hat{E}_\tau^{a}\; , \qquad
\hat{E}_\tau^{a}= \partial_\tau \hat{Z}^M E_M^a(\hat{Z}^M(\tau))\; .  \quad
  \end{eqnarray}

The supervielbein  $E^a= dZ^ME_M^a(Z)$ should obey the 11D superspace supergravity constraints \cite{BrinkHowe80}. In this paper we will mainly restrict ourselves by the case of flat target superspace for which
\begin{eqnarray}
\label{Ea=Pi} E^a = \Pi^a = dx^a - i d\theta \Gamma^a\theta\; , \quad E^\alpha=d\theta^\alpha
 \; .  \quad
  \end{eqnarray}
Here and below we use the real 32$\times$32 matrices   $\Gamma^a_{\alpha\beta}=\Gamma^a_{\beta\alpha}= \Gamma^a_{\alpha}{}^\gamma C_{\gamma\beta}$ and
$\tilde{\Gamma}_a^{\alpha\beta}=\tilde{\Gamma}_a^{\beta\alpha}= C^{\alpha\gamma}\Gamma^a_{\gamma}{}^{\beta}$  constructed as a product of the 11D Dirac matrices $\Gamma^a{}_{\beta}{}^\gamma$ (pure imaginary in our mostly  minus conventions $\eta^{ab}=diag (1,-1, ..., -1)$ and obeying $\Gamma^a\Gamma^b + \Gamma^b\Gamma^a=\eta^{ab} I_{32\times 32}$) with
11D charge conjugation matrix  $C_{\gamma\beta}=- C_{\beta\gamma}$  and its inverse
$C^{\alpha\beta}=-C^{\beta\alpha}$.

The action (\ref{S'M0=}) is invariant under the local fermionic $\kappa$--symmetry
 \cite{kappaAL,kappaS}
\begin{eqnarray}
\label{kappa}
\delta_\kappa \hat{x}^a =  - i  \hat{\theta} \Gamma^a\delta_\kappa\hat{\theta}\; , \qquad
\delta_\kappa \hat{\theta}^\alpha=p_a \tilde{\Gamma}^{a\alpha\beta}\kappa_\beta (\tau)\; , \qquad \delta_\kappa e=4i
\kappa_\beta \partial_\tau \hat{\theta}^\beta
 \; .  \qquad
  \end{eqnarray}
This symmetry   is important because it reflects the supersymmetry preserved by ground state of the dynamical system (thus insuring that it is a ${1\over 2}$BPS state). On the other hand, as it was discovered in \cite{stv},
it can be identified with the worldline supersymmetry of the superparticle. This fact is not manifest due to the infinite reducibility of the $\kappa$--symmetry \cite{kappaS} (which results in that the 32 parameters in the 11D Mayorana spinor  $\kappa_\alpha(\tau)$ can be used to remove only $16$ component of the fermionic
coordinate function $\hat{\theta}^\alpha(\tau)$).

\section{M0--brane in spinor moving frame formulation}

The $\kappa$--symmetry appears in its irreducible form in the so--called  spinor moving frame formulation of superparticle  \cite{B90,IB+AN=95,IB07:M0}.
The spinor moving frame action of M$0$--brane reads
\begin{eqnarray}
\label{SM0=} S_{M0} &=& \int_{W^1} \rho^{\#}\, \hat{E}^{=} = \int_{W^1} \rho^{\#}\,
u_a^{=}  \, {E}^a(\hat{Z}) \qquad
\\ \label{SM0==}
& =& {1\over 16}\int_{W^1}\rho^{\#}\, (v_{q}^{\; -}{\Gamma}_a v_{q}^{\; -}) \,
\hat{E}^{a}  \; ,
  \end{eqnarray}
where $\rho^{\#}(\tau)$  is a Lagrange multiplier and $u_a^{=}=u_a^{=}(\tau)$ is a light--like 11D vector,  $u^{=a}u_a^{=}=0$. This can be considered as a kind of square of any of the $16$ spinors $v^{-\alpha}_q=v^{-\alpha}_q(\tau)$
(which have appeared in (\ref{SM0==})) provided these are constrained by
\begin{eqnarray} \label{Iu--=vGv}
\left. \matrix{ v_q^{-\alpha} (\Gamma^a)_{\alpha\beta} v_p^{-\beta}= \delta_{qp}
u^{=}_{ a} \;  \cr 2v_q^{-\alpha} {}v_q^{-\beta} =
u^{=}_{ a}  \tilde{\Gamma}^{a\alpha\beta} \;  }\right\}
\quad  \Rightarrow \quad u^{=a}u_a^{=}=0 \; .
\end{eqnarray}
With the use of these constrained spinors, the $\kappa$--symmetry of the spinor moving frame action can be written in the following irreducible form
\begin{eqnarray}
\label{kap=irr}
\delta_\kappa \hat{x}^a =  - i  \hat{\theta} \Gamma^a\delta_\kappa\hat{\theta}\; , \qquad
\delta_\kappa \hat{\theta}^\alpha = \epsilon^{+q} (\tau)  v_q^{-\alpha} \; , \qquad  \delta_\kappa \rho^{\#}=0= \delta_\kappa u_a^{=}
 \; .  \qquad
\end{eqnarray}
We have denoted the parameter of the irreducible  $\kappa$--symmetry by $ \epsilon^{+q} = \epsilon^{+q} (\tau)$ to stress its relation with the extended worldline supersymmetry.

The transformations (\ref{kap=irr}) can be obtained from the infinitely reducible (\ref{kappa}) by substituting for $p_a$ the solution $p_a=\rho^{\#}u_a^=$ of the  constraint $p_ap^a=0$. It is easy to see that,  with this substitution, the action (\ref{S'M0=}) acquires the form of (\ref{SM0=}). Furthermore, using (\ref{Iu--=vGv}), we find  \begin{eqnarray}
\label{epq=vkap}\epsilon^{+q}= 2 \rho^{\#}v_q^{-\alpha}\kappa_\alpha\; .  \qquad
\end{eqnarray}

However, one might still find the origin of our $v^{-\alpha}_q$ a bit mysterious. To clarify this, it is useful to consider the  null--vector $u_a^{=}$ as an element of the
 {\it moving frame} matrix,
 \begin{eqnarray}\label{Uin}
& U_b^{(a)}= \left({u_b^{=}+ u_b^{\#}\over 2}, u_b^{i}, {
u_b^{\#}-u_b^{=}\over 2} \right)\; \in \; SO(1,10)\;  \quad
\end{eqnarray}
($i=1,...,9$). The statement that this matrix belongs to the $SO(1,10)$ is tantamount to saying that the moving frame vectors obey the constraints \cite{Sok}
\begin{eqnarray}\label{u--u--=0}
u_{ {a}}^{=} u^{ {a}\; =}=0\; , \quad    u_{ {a}}^{\; = } u^{ {a} \#}= 2\; , \quad u_{ {a}}^{=} u^{ {a}\,i}=0\; , \qquad
 \\  \label{u++u++=0} u_{ {a}}^{\# } u^{ {a} \#
}=0 \; , \quad
 u_{{a}}^{\;\#} u^{ {a} i}=0\; , \qquad  \\ \label{uiuj=-} u_{ {a}}^{ i} u^{{a}j}=-\delta^{ij}\; .  \quad
\end{eqnarray}
Then $v^{-\alpha}_q$ can be defined as an 16$\times$32 block of  $Spin(1,10)$ valued {\it spinor moving frame} matrix  \begin{eqnarray}\label{harmVin}
V_{(\beta)}^{\;\;\; \alpha}= \left(\matrix{  v^{+\alpha}_q
 \cr  v^{-\alpha}_q } \right) \in Spin(1,10)\;
 \; . \qquad
\end{eqnarray}
This is double covering of the moving frame matrix (\ref{Uin}) in the sense of that
 \begin{eqnarray}\label{VGVt=G} V\Gamma_b V^T =  U_b^{(a)} {\Gamma}_{(a)}\, , \quad (a) \qquad V^T \tilde{\Gamma}^{(a)}  V = \tilde{\Gamma}^{b} u_b^{(a)}\;
 \, , \quad (b) \qquad   VCV^T=C \; . \quad (c) \qquad
\end{eqnarray}
 The two seemingly mysterious constraints (\ref{Iu--=vGv}) appear as a block of the relation (\ref{VGVt=G}a) and as the $=$ component,   $V^T \tilde{\Gamma}^{=}  V = \tilde{\Gamma}^{b} u_b^{=}$, of (\ref{VGVt=G}b),
with an appropriate representation of the 11D Gamma matrices. The other blocks/components of these constraints involve the second set of constrained spinors, $v_{q}^{+ {\alpha}}$,
\begin{eqnarray}\label{M0:v+v+=u++}
 v_{q}^+ {\Gamma}_{ {a}} v_{p}^+ = \; u_{ {a}}^{\# } \delta_{qp}\; , \qquad
 v_{q}^- {\Gamma}_{ {a}} v_{p}^+ = - u_{ {a}}^{i} \gamma^i_{qp}\; , \qquad
\\ \label{M0:u++G=v+v+}
 2 v_{q}^{+ {\alpha}}v_{q}^{+}{}^{ {\beta}}= \tilde{\Gamma}^{ {a} {\alpha} {\beta}} u_{ {a}}^{\# }\; , \quad
 2 v_{q}^{-( {\alpha}}v_{q}^{+}{}^{ {\beta})}=-  \tilde{\Gamma}^{ {a} {\alpha} {\beta}} u_{ {a}}^{i}\; . \quad
\end{eqnarray}
Here $\gamma^i_{qp}$ are the 9d Dirac matrices; they are real, symmetric  $\gamma^i_{qp}=\gamma^i_{pq}$ and obey $\gamma^i\gamma^j + \gamma^j \gamma^i= 2\delta^{ij} I_{16\times 16}$.

The third constraint, (\ref{VGVt=G}c), implies that the inverse spinor moving frame matrix
\begin{eqnarray}\label{Vharm=M0}
 V^{( {\beta})}_{ {\alpha}}= \left(
v_{ {\alpha}q}{}^+\, ,v_{ {\alpha}q}{}^- \right)\;
\in \; Spin(1,10)  \qquad
\end{eqnarray}
can be constructed from the same  $ v_{q}^{\mp {\alpha}}$ as in (\ref{harmVin}),
\begin{eqnarray}
\label{V-1=CV}  v_{\alpha}{}^{\mp}_q = \pm i
C_{\alpha\beta}v_{q}^{\mp \beta }\quad \Rightarrow \quad \cases{ v_{q}^{- {\alpha}}v_{ {\alpha}p}{}^+=\delta_{qp}= v_{q}^{+ {\alpha}}v_{ {\alpha}p}{}^-\, , \cr  v_{q}^{- {\alpha}}v_{ {\alpha}p}{}^-= 0\; = v_{q}^{+ {\alpha}}v_{ {\alpha}p}{}^+\, .
}
 \end{eqnarray}

The moving frame vectors can be used to split the pull--back of the supervielbein in a Lorentz covariant manner, \begin{eqnarray}
\label{Ea->EUa}\hat{E}^b\mapsto \hat{E}^b U_b^{(a)}= ( \hat{E}^{=}, \hat{E}^{\#}, \hat{E}^i)\; . \qquad \end{eqnarray} One can show that the equations of motion for the Lagrange multiplier $\rho^{\#}$ and for the moving frame vector $u_a^{=}$ (or for $v_q^{-\alpha}$) result in $\hat{E}^{=}=0$ and $\hat{E}^i=0$, respectively (see \cite{IB07:M0,IB09:M-D,mM0=PLB} for details on varying the moving frame and spinor moving frame fields), so that on the mass shell
\begin{eqnarray}\label{E==0=Ei}
\left. \matrix{ \hat{E}^{=}:= \hat{E}^a u_{ {a}}^{=}=0\;  \cr   \hat{E}^{i}:= \hat{E}^a u_{ {a}}^{i}=0\;  }\right\} \quad   \Leftrightarrow \qquad \hat{E}^{a}:= {1\over 2}\; \hat{E}^{\#} u_{ {a}}^{=} \; .  \qquad
\end{eqnarray}
This implies that the M$0$--brane worldline $W^1$ is a light--like line in target (super)space, which is in agreement with the statement that M$0$--brane is the massless 11D superparticle.

Furthermore (\ref{E==0=Ei}) suggests to consider $\hat{E}^{\#}=d\tau \hat{E}_\tau^{\#}$ as an einbein on $W^1$. Its gravitino--like  companion is given by the covariant projections $ \hat{E}^{+q}=  \hat{E}^{\alpha}v_\alpha^{+q}$
of the pull--back of the fermionic 1--form $E^\alpha$. One can show \cite{IB07:M0} that the other projection,  $\hat{E}^{-q}=  \hat{E}^{\alpha}v_\alpha^{-q}$, vanishes due to the fermionic equation of motion of the M$0$-brane, so that, on the mass shell,
\begin{eqnarray}\label{E-q=0}
\hat{E}^{-q}:=  \hat{E}^{\alpha}v_\alpha^{-q}=0  \qquad   \Leftrightarrow \qquad \hat{E}^{\alpha}:= \hat{E}^{+q} v_q^{-\alpha} \; .  \qquad
\end{eqnarray}
The suggestion to treat $ \hat{E}^{+q}=  \hat{E}^{\alpha}v_\alpha^{+q}=d\tau \hat{E}{}^{+q}_{{\tau}}$  as composed gravitino and  $(\hat{E}_\tau^{\#},\hat{E}{}^{\;+q}_{{\tau}})$ as composed supergravity multiplet  induced by embedding of $W^1$ into the 11D target superspace may be taken  from the observation that under the irreducible $\kappa$--symmetry (\ref{kap=irr})
\begin{eqnarray}\label{v1dSG=}
\delta_\kappa \hat{E}^{+q}= D \epsilon^{+q}(\tau) \; ,  \qquad \delta_\kappa \hat{E}^{\#}= -2i \hat{E}^{+q}\epsilon^{+q}\;
\end{eqnarray}
($D=d\tau D_\tau$ is defined below, in (\ref{DPsii=})). Our  action for the mM$0$ system, which we are going to present, contains the coupling of these induced 1d supergravity to the matter describing the relative motion of the mM$0$ constituents.

\section{Covariant action for multiple M0--brane (mM0) system}

The study of \cite{mM0=PLB,mM0=PRL}  suggests that,  describing the system of
$N$ nearly coincident  M$0$-branes (mM$0$ system), it is convenient to separate the   coordinate functions $\hat{Z}{}^M(\tau)$  describing the center of energy motion (with the same properties as the ones describing single M$0$--brane) and the variables describing the relative motion of the mM$0$ constituents. That are the bosonic and fermionic hermitian traceless $N\times N$ matrix fields   ${\bb X}^i(\tau)$ and  $\Psi_q (\tau)$ depending on a  proper time variable $\tau$ parametrizing the center of energy worldline $W^1$.
The bosonic ${\bb  X}^i(\tau)$ carries the index  $i=1,...,9$ of the vector representation of $SO(9)$, while the fermionic $\Psi_q$ transforms as a spinor under $SO(9)$, so that  $q=1,...,16$. The $SO(1,1)$ weight of the fields are 2 and 3, respectively, so that in a more explicit notation ${\bb  X}^i= {\bb  X}_{\#}^i:= {\bb  X}_{++}^i$ and   ${{ \Psi}}_q= \Psi_{\# \,+q}:= \Psi_{++ \,+q}= \Psi_{\#}{}_q^-$.

We propose to describe the system of $N$ nearly coincident M$0$--branes by the following action
\begin{eqnarray}
\label{SmM0=}  S_{mM0} &=& \int_{W^1} \rho\, \hat{E}^{=} +  \int_{W^1} \rho^3\, \left(  tr\left( -{\bb P}^i D {\bb X}^i + 4i { \Psi}_q D { \Psi}_q  \right) + \hat{E}^{\#} {\cal H} \right) + \quad \nonumber \\
&&  +  \int_{W^1} \rho^3\, \hat{E}^{+q}  tr\left(4i (\gamma^i {\Psi})_q  {\bb P}^i + {1\over 2} (\gamma^{ij} {\Psi})_q  [{\bb X}^i, {\bb X}^j]  \right)\; .
  \end{eqnarray}
In it  the measure factor $d\tau$ is hidden in $D=d\tau D_\tau$ (described below) and inside the bosonic and fermionic one forms\footnote{Let us recall that we consider the case of flat target 11D superspace, in which  $\hat{E}^a= d\tau
(\partial_\tau \hat{x}^a - i \partial_\tau \hat{\theta}\Gamma^a\hat{\theta})$,
and $ \hat{E}^\alpha =d\hat{\theta}^\alpha= d\tau
\, \partial_\tau  \hat{\theta}^\alpha$, see Eqs.  (\ref{hEa=dthEt}) and (\ref{Ea=Pi}).}
\begin{eqnarray}
\label{hE==hEu=} \hat{E}^{=}=
\hat{E}^{a}u_a^=  =d\tau \hat{E}^{=}_\tau
\, , \qquad \hat{E}^{\#}=\hat{E}^au_a^{\#}=d\tau \hat{E}^{\#}_\tau  \; ,  \qquad    \label{hE+q=hEv+q}
\hat{E}^{+q}=\hat{E}^\alpha v_\alpha^{+q}= d\tau \hat{E}^{+q}_\tau \; . \qquad
  \end{eqnarray}
These, as well as the Lagrange multiplier $\rho=\rho^{\#}(\tau)$, have been described above, in sec. {\bf 2}, when discussing the single mM$0$-brane case. But now the moving frame vectors $u_a^=$ and $u_a^{\#}$, obeying (\ref{u--u--=0}) and (\ref{u++u++=0}),   and spinor moving frame variable $v_\alpha^{+q}$, obeying (\ref{M0:v+v+=u++}) and (\ref{M0:u++G=v+v+}), are related to the center of energy motion of the mM$0$ system. Further, ${\bb P}^i:={\bb P}_{\#\#}^i$ are 9 auxiliary  bosonic matrix  fields having the meaning of the momentum of ${\bb X}^i$ and
\begin{eqnarray}
\label{HmM0=} {\cal H} &:= & {\cal H}_{\#\#\#\#}({\bb X}, {\bb P}, \Psi )  = {1\over 2} tr\left( {\bb P}^i {\bb P}^i \right) + {\cal V} ({\bb X}) - 2  tr\left({\bb X}^i\, \Psi\gamma^i {\Psi}\right)  \qquad
  \end{eqnarray}
is the relative motion Hamiltonian. Besides  the kinetic term $tr({\bb P}^i)^2$, this includes  the  Yukawa coupling  $tr\left({\bb X}^i\, \Psi\gamma^i {\Psi}\right)$ and
the scalar  potential
\begin{eqnarray}
\label{VmM0=}  {\cal V} := {\cal V}_{\#\#\#\#} ({\bb X}) = -  {1\over 64} tr\left[
{\bb X}^i ,{\bb X}^j \right]^2 \; . \qquad
  \end{eqnarray}
The covariant derivatives $D=d\tau D_\tau $ are defined by
\begin{eqnarray}
\label{DXi=} D{\bb X}^i  &:=& d{\bb X}^i  + 2\Omega^{(0)} {\bb X}^i  - \Omega^{ij} {\bb X}^j+ [{\bb A},    {\bb X}^i] \; , \qquad \\
\label{DPsii=} D\Psi_q  &:=& d\Psi_q
 + 3\Omega^{(0)} \Psi_q   -{1\over 4} \Omega^{ij} \gamma^{ij}_{qp} {\Psi}_p+ [{\bb A},   \Psi_q ] \; . \qquad
  \end{eqnarray}
Here ${\bb A}= d\tau {\bb A}_\tau (\tau)$ is the $SU(N)$ connection on $W^1$,
 ${\bb A}_\tau (\tau)$ is an anti-Hermitian traceless $N\times N$ matrix gauge field in 1d, which is an independent variable in our model. In contrast,  $\Omega^{(0)}=d\tau \Omega_\tau^{(0)}$ and $\Omega^{ij}=d\tau \Omega_\tau^{ij}$ are the composed (induced) $SO(1,1)$ and $SO(9)$ connections on $W^1$. They are constructed from the moving frame vector fields (\ref{Uin}) corresponding to the center of energy motion as
\begin{eqnarray}
\label{Om0=} \Omega^{(0)}= {1\over 4} u^{=a}du_a^{\#}\; , \qquad \Omega^{ij}=  u^{ia}du_a^{j}\; . \qquad
  \end{eqnarray}

The action (\ref{SmM0=}) is invariant under the transformations of the ${\cal N}=16$ local worldline supersymmetry `parameterized' by fermionic $SO(9)$ spinor function $\epsilon^{+q}=\epsilon^{+q} (\tau)$. This acts on the matrix fields describing the relative motion of the mM$0$ constituents as\footnote{To prove the invariance of the action (\ref{SmM0=}) under these supersymmetry transformations the following identity for 9d gamma matrices is useful:
$\gamma^{ij}_{q(q^\prime }\gamma^{i}_{p^\prime)p }+ \gamma^{ij}_{p(q^\prime }\gamma^{i}_{p^\prime)q } = \gamma^{j}_{q^\prime p^\prime}\delta_{qp}-\delta_{q^\prime p^\prime}\gamma^{j}_{qp} $.}.
\begin{eqnarray}
\label{susy-X}
\delta_\epsilon {\bb X}^i   = 4i \epsilon^{+} \gamma^i  \Psi \; , \qquad \delta_\epsilon {\bb P}^i   = [(\epsilon^{+} \gamma^{ij}  \Psi),  {\bb X}^j]\; ,\qquad \\
\label{susy-Psi}
\delta_\epsilon \Psi_q =  {1\over 2} (\epsilon^{+} \gamma^i)_q  {\bb P}^i-  {i\over 16} (\epsilon^{+} \gamma^{ij})_q  [{\bb X}^i, {\bb X}^j]\; ,\qquad \\ \label{susy-A}
 \delta_\epsilon {\bb A} = -  \hat{E}^{\#} \epsilon^{+q}  \Psi_q + (\hat{E}^{+}\gamma^i \epsilon^{+})    {\bb X}^i
 \; ,  \qquad
\end{eqnarray}
and on the center of energy variables as
\begin{eqnarray}
\label{susy-x}
\delta_\epsilon \hat{x}^a
&=& - i \hat{\theta} \Gamma^a\delta_\epsilon \hat{\theta}+ 3 \rho^2 u^{a\#}tr \left( i(\epsilon^{+} \gamma^{i}\Psi) {\bb P}^i -  (\epsilon^{+} \gamma^{ij}\Psi) [{\bb X}^i, {\bb X}^j]/8    \right)
\; , \qquad \\
 \label{susy-th}
\delta_\epsilon \hat{\theta}^\alpha &=& \epsilon^{+q}  v_q^{-\alpha} \; , \qquad
 \\
 \label{susy-u}   && \delta_\epsilon \rho =0= \delta_\epsilon u_a^{=}
 \; .   \qquad
\end{eqnarray}
Eqs. (\ref{susy-x}), (\ref{susy-th}) and (\ref{susy-u}) describe a deformation of the  irreducible $\kappa$--symmetry (\ref{kap=irr}) of the free massless superparticle. Actually the only deformed relation is $\delta_\epsilon \hat{x}^a$, (\ref{susy-x}),  which acquires an additional (with respect to  (\ref{kap=irr})) contribution  constructed from ${\bb X}^i$, ${\bb P}^i$ and $\Psi_q$.

The  local supersymmetry (\ref{susy-X})-- (\ref{susy-u}) guaranties that the ground state of the dynamical system described by the action  (\ref{SmM0=}) preserves 1/2 of 32 11D supersymmetries (is a 1/2 BPS state), the fact which  allows to identify (\ref{SmM0=}) with the action of multiple M$0$-brane system.

\section{Effective mass of the center of energy motion of the mM0 system}

The fact that the  worldline supersymmetry transformations of the mM0 center of energy variables are so close to the $\kappa$--symmetry transformations of single M$0$-brane can be traced to the fact that  the first term in (\ref{SmM0=}) coincides with the action (\ref{SM0=}) of the single  M$0$--brane. However, due to the presence of the Lagrange multiplier  $\rho (\tau) =\rho^{\#}(\tau)$ and of the  moving frame variables, $u_a^{\#}$ and $v_\alpha^{+q}$, also in
the second part of the action (containing $\hat{E}^{\#}= \hat{E}^au_a^{\#}$ and $\hat{E}^{+q}= \hat{E}^\alpha v_\alpha^{+q}$), the equations for the auxiliary fields differ  from (\ref{E==0=Ei}) in such a way that, in contrast with the case of a single M$0$-brane, generic motion of the center of energy  of the mM0 system is not light-like; it is characterized by a nonvanishing effective mass constructed from the relative motion fields ${\bb X}^i$ and $\Psi_q$.

\subsection{Effective mass of the center of energy motion}

To see this in a simple way, let us calculate the canonical momentum conjugate to the center of energy coordinate function $\hat{x}^a(\tau)$. This reads
\begin{eqnarray}
\label{pa=}
p_a(\tau)= {\partial {\cal L}_{mM0}\over \partial \partial_\tau \hat{x}^a(\tau) }=
\rho u_a^= + (\rho)^3 u_a^\# {\cal H}({\bb X}^i, {\bb P}^i,\Psi_q)
\; , \qquad
  \end{eqnarray}
where ${\cal L}_{mM0}$ is the Lagrangian density of the action (\ref{SmM0=}),
$S_{mM0}=\int d\tau {\cal L}_{mM0}$, and
${\cal H}={\cal H}_{\#\#\#\#}$ is defined in Eq. (\ref{HmM0=}).
Now, using Eqs. (\ref{u--u--=0}) and (\ref{u++u++=0}) one easily finds that
\begin{eqnarray}
\label{M2=11D}
M^2:= p^ap_a(\tau)=4
\rho^4  {\cal H}({\bb X}^i, {\bb P}^i,\Psi_q)
\; . \qquad
  \end{eqnarray}

In purely bosonic limit it is easy to see that $M^2$ is a
nonnegative constant. Indeed, when $\Psi=0$ the relevant equations of the relative motion which follow from the action (\ref{SmM0=}) read
\begin{eqnarray}
\label{Pi=DXi}
&& \hat{E}^{\#} {\bb P}^i= D{\bb X}^i \quad \Rightarrow \quad  {\bb P}^i= D_{\#} {\bb X}^i  \; , \qquad
 \\ \label{DDX=XXX} && D_{\#} D_{\#} {\bb X}^i={1\over 16} [[{\bb X}^i, {\bb X}^j]{\bb X}^j] \; ,
  \end{eqnarray}
where $D_{\#} $ is defined by $D_{\#} := D_{\tau}/ \hat{E}_\tau^{\#}$ or, equivalently,  by
$D= E^{\#} D_{\#}= d\tau  D_{\tau}$ where $D$ is the covariant derivative defined in (\ref{DXi=}), (\ref{DPsii=}). As a consequence of Eqs. (\ref{Pi=DXi}) and (\ref{DDX=XXX}), the relative motion Hamiltonian is covariantly constant, $D{\cal H}=0$. As far as ${\cal H}={\cal H}_{\#\#\#\#}$ carries the weight 8 with respect to SO(1,1), the covariant derivative  involves the induced SO(1,1) connection (\ref{Om0=}) so that a more explicit form of this equation is $d{\cal H}+ 8\Omega^{(0)}{\cal H}=0$. Now we shall notice that the set of equations of motion following from the action (\ref{SmM0=}) also includes $D\rho:=d\rho - 2  \Omega^{(0)}\rho =0$  \footnote{See \cite{IB07:M0} and forthcoming \cite{IB+CM=mM0} for details on variation of 11D spinor moving frame variables; in this paper we would like to omit such type technicalities.  Remember that $\rho =\rho^{\#}(\tau)$ has SO(1,1) weight -2, opposite to the weight of  ${\bb X}^i= {\bb X}^i_\#(\tau)$.} which can be solved for $SO(1,1)$ connection $\Omega^{(0)}= {d\rho/2 \rho}$. Using this solution we see that  $D{\cal H}=0$ can be written in the form of ${d(\rho^4{\cal H})\over \rho^4} =0$ which makes manifest that  $M^2$ in (\ref{M2=11D}) is a constant. This constant is nonnegative just due to the explicit form of (the bosonic limit of) the relative motion Hamiltonian  ${\cal H}$, Eqs. (\ref{HmM0=}), (\ref{VmM0=}).

\subsection{On relation with the results of \cite{mM0=PLB,mM0=PRL}}
\label{relation}

This is the place to comment on relation of our model with \cite{mM0=PLB,mM0=PRL}. The reader who is not interested  in superembedding approach can omit this subsection and pass directly to the next sec. \ref{BPS:M2=0}.

The complete equations of motion following from the action (\ref{SmM0=})  are close to, but not identical with, the flat superspace case of the mM0 equations  obtained in  \cite{mM0=PLB,mM0=PRL} in the frame of the superembedding approach. The difference is due to the  contribution of the center of energy fields into the equations of relative motion and vice versa (see \cite{IB+CM=mM0} for details).  For instance, the complete form of the the bosonic equation of the relative motion which follows from the action (\ref{SmM0=}) reads
\begin{eqnarray}
\label{D2Xi=}
D_{\#} D_{\#} {\bb X}^i={1\over 16} [[{\bb X}^i, {\bb X}^j]{\bb X}^j] + 2i \Psi\gamma^i\Psi + 4i D_{\#} (\hat{E}_\#^{+}\gamma^i\Psi) +i (\hat{E}_\#^{+}\gamma^{ij})_q[{\Psi}_q, {\bb X}^j] \; .
  \end{eqnarray}
The last two terms in the r.h.s. of (\ref{D2Xi=}), which contain contributions of the center of energy Goldstone fermion $\hat{\theta}^\alpha$ (through $\hat{E}_\#^{+q} := D_\# \hat{\theta}^\alpha  v_\alpha^{+q} =  \partial_\tau \hat{\theta}^\alpha  v_\alpha^{+q} /\hat{E}_\tau^\#$) are absent, in the flat target superspace case of equations from \cite{mM0=PLB,mM0=PRL}. The reason for such a difference is that the center of energy motion is considered  as a (worldline superspace) background in  \cite{mM0=PLB,mM0=PRL} while in our present action (\ref{SmM0=}) it is treated on the same footing as the relative motion.

Due to the same reason, the approach of \cite{mM0=PLB,mM0=PRL} did not catch  the backreaction of the relative motion on the center of energy motion (which is also characteristic for the purely bosonic Myers action \cite{Myers:1999ps}), one of the manifestation of which is the appearance of generically nonvanishing  effective mass of the mM0 system, Eq. (\ref{M2=11D}), which we discussed above in the frame of purely bosonic approximation.

To make the approach of \cite{mM0=PLB,mM0=PRL} accounting for the mutual influence of the center of energy and relative motion, the basic center of energy superembedding equation, which was taken in \cite{mM0=PLB,mM0=PRL} to be the same as in the single M$0$ case, should be modified by the terms constructed with the use of the relative motion superfields. In the next section we will comment on one of the possible ways to find such a modification.

\subsection{$M^2=0$ as a BPS equation. Vanishing effective mass of all supersymmetric bosonic solutions}
\label{BPS:M2=0}

Interestingly enough, all the supersymmetric bosonic solutions of the equations which follow from the action (\ref{SmM0=}) are characterized by the vanishing effective mass, $M^2=0$. Indeed, setting $\Psi_q=0$ we find from (\ref{susy-Psi}) the following Killing spinor equation
\begin{eqnarray}
\label{Killing}
  \epsilon^{+p}{\bb K}_{pq} := (\epsilon^{+} \gamma^i)_q  {\bb P}^i-  {i\over 8} (\epsilon^{+} \gamma^{ij})_q  [{\bb X}^i, {\bb X}^j]=0\; .\qquad
\end{eqnarray}
Its consistency condition $\epsilon^{+p}  tr ( ({\bb K}\gamma^j)_{pq}{\bb P}^j)  + {i\over 8}({\bb K}\gamma^{jk})_{pq}[{\bb X}^j, {\bb X}^k])=0$ can be presented in the form $\epsilon^{+q} {\cal H}=0$ \footnote{  To this end one have to use Jacobi identities for commutators and the properties of 9d gamma matrices.}. Taking into account Eq. (\ref{M2=11D}), this can be written as $\epsilon^{+q} M^2=0$. Hence (one of) the BPS equation(s) for  supersymmetric bosonic solutions of mM0 equations is
\begin{eqnarray}
\label{BPS}
M^2=0 \qquad \Leftrightarrow \qquad {\cal H}\vert_{\Psi=0}={1\over 2} tr\left( {\bb P}^i {\bb P}^i \right)-  {1\over 64} tr\left[
{\bb X}^i ,{\bb X}^j \right]^2=0\;  . \qquad
\end{eqnarray}

This fact is very important. It implies that a supersymmetric solution of the 11D supergravity describing  our mM$0$ system has the same property as the single M$0$--brane (M-wave) solution. In other words, our result does not imply the existence of a new exotic supersymmetric solution of the 11D supergravity.

Furthermore, the form of the Hamiltonian of the relative motion ${\cal H}$, which essentially coincides with $M^2$, Eq. (\ref{M2=11D}),  indicates that   all supersymmetric bosonic solutions have ${\bb P}^i=0$ and $ \left[{\bb X}^i ,{\bb X}^j \right]=0$, {\it i.e.} that their relative motion sector  is in its  ground state.

\section{Conclusion and discussion}

In this paper we have presented the complete action for the multiple M$0$--brane (multiple M-waves or mM$0$) system  in flat target 11D superspace. This action is Lorentz covariant, invariant under the global 11D target space supersymmetry and under the local 1d ${\cal N}=16$  supersymmetry which acts on the center of energy variables like a deformed version of the $\kappa$--symmetry of the massless superparticle. We have also shown that the generic motion of the mM$0$ system is characterized by a nonnegative center of energy mass constructed from the relative motion variables so that the center of energy worldline is generically not light--like. On the other hand, we have shown that all the supersymmetric bosonic solutions of mM$0$ equations  are characterized by vanishing effective mass, $M^2=0$.

The importance of this observation  is related to the fact that supersymmetric extended objects, for instance M2 and M5-branes, can be described not only by worldvolume actions, but also by supersymmetric  solutions of the appropriate (in this case 11D) supergravity equations. Indeed, if it were found that there existed a supersymmetric solution of mM$0$ equations (mM$0$ BPS state) with $M^2\not=0$, this would imply the existence  of new exotic supersymmetric solution of the 11D supergravity equations. In contrast, our result that all the supersymmetric mM$0$ BPS states are massless, $M^2=0$, imply that the 11D supergravity solutions describing them are similar to a simple M-wave describing the single M$0$-brane (see \cite{TomasBook} for discussion on this solution).

The detailed study of the equations of motion which follow from the multiple M-wave action
(\ref{SmM0=}) and search for their solutions will be the subject of the forthcoming paper
\cite{IB+CM=mM0}. Here we restrict ourself by commenting on the relation with \cite{mM0=PLB,mM0=PRL} in sec. \ref{relation} and on relation with \cite{Dima+=JHEP03} below.

The dimensional reduction of our action on a circle  should be related to (the moving frame reformulation of) the mD$0$ action from \cite{Dima+=JHEP03}. The generalized mass term of the 10D  model of \cite{Dima+=JHEP03} is defined by an arbitrary function $M_{10D}= M_{10D}(\tilde{\bb P}^i, \tilde{\bb X}^i, \tilde\Psi)$, while the dimensional reduction of our action gives more definite expression,
\begin{eqnarray}\label{M2D10=}
M^2_{10D}=p_0^2-p_1^2- ...- p^2_9= p_{10}^2 + 4 \rho^4 {\cal H}\; . \qquad
\end{eqnarray}
Here  ${\cal H}= {\cal H}_{\#\#\#\#} ({\bb P}^i, {\bb X}^i, \Psi_q)$ is the  Hamiltonian of the multiple M$0$ system, Eq. (\ref{HmM0=}), and,  as we discussed above, $4 \rho^4\, {\cal H}=M^2$ is a constant 11D effective mass of the mM$0$ system.
Notice however that in our expression for $M_{10D}^2$  some freedom is still present: we have to  choose the form of the momentum  $p_{10}$ corresponding to the compactified direction. Thus the possibility to reproduce the counterpart of the mD0 model from \cite{Dima+=JHEP03} starting from our mM$0$ action is related with an exotic dimensional reduction defined with the use of the relative motion variables.

The study of dimensional reductions and the search for possible reformulation of our model without spinor moving frame variables are interesting problems for future study. On the other hand, we intend to use our spinor moving frame action as a basis of the generalized action principle \cite{bsv} and to study the superfield equations which can be derived from it. This will produce such a deformation of the basic superembedding equation that will result in a modification of  the dynamical  equations from  \cite{mM0=PRL,mM0=PRD} by terms describing the influence of the relative motion fields on the center of energy motion and {\it vice versa}.

The other impotent direction for future study is to elaborate the generalization of the action (\ref{SmM0=}) for the mM$0$ system in curved 11D superspace\footnote{Notice that in our case the problem of coupling to 11D supergravity seems to be pure technical (although  probably difficult). This differs from the problem of whether the Matrix model provides the description of the complete M-theory in some particular frame, discussed in \cite{D+O=97}. }. The form of equations in \cite{mM0=PRL} suggests that, besides  understanding the ${E}^a$ and ${E}^\alpha$, included in (\ref{SmM0=}) inside of $\hat{E}^{\#}$, $\hat{E}^{=}$ and $\hat{E}^{+q}$, to be  supervielbein of the 11D supergravity superspace (instead of  (\ref{Ea=Pi})), one should also add an explicit interaction with the field strengths (''fluxes'') of the 11D supergravity, $F_{abcd}= F_{[abcd]}(Z)$, $R_{abcd}(Z)$ and $T_{ab}^\alpha(Z)$.
The terms suggested by the equations from   \cite{mM0=PRL,mM0=PRD} read $\Delta^{fluxes} S_{mM0}=  \int_{W^1}  \hat{E}^{a} (\rho^{\#})^3\, {\cal L}_a^{fluxes}$ with
\begin{eqnarray}
\label{LmM0=} {\cal L}^{a\; fluxes} =
 {1\over 4!}\, \hat{F}^{a ijk}
 tr\left( {\bb X}^i[{\bb X}^j, {\bb X}^k ]+ 4i  {\Psi} \gamma^{ijk}{\Psi} \right)
+ {1\over 8}\; \hat{R}^{a\, i=j} tr\left( {\bb X}^i{\bb X}^j \right) +  2i \hat{T}^{a\, i-q}  tr\left({\bb X}^i {\Psi}_q  \right) \, ,
\;
  \end{eqnarray}
and $\hat{F}^{a ijk}={F}^{abcd}(\hat{Z}) u_b^{i} u_c^{j}u_d^{k}$, $\hat{R}^{a i=j}={R}^{abcd}(\hat{Z}) u_b^{i} u_c^{=}u_d^{j}$, $\hat{T}^{a i-q}={T}^{ab\alpha}(\hat{Z}) u_b^{i} v_{\alpha}^{-q}$.
The explicit presence of the flux superfields creates some difficulties in the  calculations so that the question of  whether $ S_{mM0}{}^{[Eq. (\ref{SmM0=})]}+ \Delta^{fluxes} S_{mM0}$ gives a supersymmetric action for mM$0$ system in the supergravity superspace, or some additional terms are needed to provide the local worldline supersymmetry ({\it i.e.} $\kappa$--symmetry), is still open. It is natural to begin with the cases of mM0 in vacuum superspaces where the fluxes acquire constant values; notice that the $AdS\times S$ and $pp$-wave superspaces are of this type.

Actually, the analogy with the bosonic multiple D$p$-brane actions of \cite{Myers:1999ps,Taylor+VanRaamsdonk}
suggests to expect the background superfields to depend on matrix coordinates. To do this in a covariant manner   one must study a model involving, schematically, something like
\begin{eqnarray}\label{E=E(x+X,t+P)} E_M^a \left(\hat{Z}^N  +
\tilde{\bb X}{}^i u^{ia}E_a^N(\hat{Z}+...) +  \tilde{\Psi}_q v_q^{+\check{\alpha}} E_{\check{\alpha}}^N(\hat{Z}+...)\right)\; . \qquad
\end{eqnarray}
Although our moving frame and spinor moving frame variables seems to be useful in writing
(\ref{E=E(x+X,t+P)}) and similar expressions, to deal with them is quite a difficult problem,   see \cite{nonAbDp} for relevant studies. However, for the case of nearly coincident branes one can use the series decomposition of the background superfields near the center of energy of multiple brane system. After such a decomposition, instead of function of non-commuting coordinates, like (\ref{E=E(x+X,t+P)}), we will have the sum of polynomials in these matrix coordinates multiplied by derivatives of the background gauge (super)fields depending on the center of energy variables only. In an appropriate gauge these can be expressed in terms of the field strength (fluxes) and their covariant derivatives calculated at the center of energy 'position' in superspace. The straightforward search for curved superspace generalization of the action (\ref{SmM0=}) corresponds to the search for such a decomposition with a hope that probably this generically infinite series can be consistently truncated to a polynomial in matrix field with preservation of local supersymmetry. If this truncation does not occur, the above line might be stopped 'by hand' at some power ($\geq 4$) in a relative motion fields ${\bb X}^i$ and $\Psi^i$, thus giving a weak field approximation to the action for mM$0$ system in 11D supergravity background.

Finally, probably the most intriguing question is whether it is possible to find a generalization of our mM$0$  action for the case of multiple M$2$--brane system. We should note that, although the search for the answer for this question does not promise to be simple, in particular in the light of the recent results in \cite{Paul+Bengt}, however, neither it looks hopeless.

{\bf Acknowledgements}.
{The author is thankful to Dima Sorokin for interest to this work which was supported in part by the research grant FIS2008-1980 from the MICINN (presently MINECO) of Spain, by the Basque Government Research Group Grant ITT559-10 and by the UPV/EHU under the program UFI 11/55.}

\end{document}